\documentstyle[aps,prb,epsfig,floats]{revtex}	

\def\ea{{\it et al.}\ }

\begin{document}

\draft

\title{Multiple Andreev reflection in single atom niobium junctions}
\author{B. Ludoph$^{1}$, N. van der Post$^{1, *}$, E.N. Bratus'$^{2}$, E.V. Bezuglyi$^{2}$, V.S. Shumeiko$^{3}$, G. Wendin$^{3}$\\
and J.M. van Ruitenbeek$^{1, \dag}$}
\address{$^{1}$ Kamerlingh Onnes Laboratorium, Universiteit Leiden, Postbus 9504, 2300 RA Leiden, The Netherlands \newline
$^{2}$ B. Verkin Institute for Low Temperature Physics and Engineering, 310164 Kharkov, Ukraine \newline
$^{3}$ Department of Microelectronics and Nanoscience, Chalmers University of Technology\\ and G\"oteborg University, SE-41296 G\"oteborg, Sweden}

\date{\today}

\maketitle

\begin{abstract}
Single atom junctions between superconducting niobium leads are produced using the Mechanically Controllable Break Junction technique. The current-voltage characteristics of these junctions are analysed using an exact formulation for a superconducting quantum point contact. For tunnelling between two single atoms with a sufficiently large vacuum barrier, it is found that a single channel dominates the current, and that the current-voltage characteristic is described by the theory, without adjustable parameters. For a contact of a single Nb atom it is shown that five conductance channels contribute to the conductance, in agreement with the number expected based on the number of valence orbitals for this {\it d}-metal. For each of the channels the transmission probability is obtained from the fits and the limits of accuracy for these numbers are discussed. 
\end{abstract}



\section{Introduction}
The current-voltage (IV) characteristic for superconducting junctions provides detailed information on the material and junction parameters. For standard planar tunnel junctions, consisting of two identical superconducting films separated by an insulating layer, the most prominent feature is the sudden rise of the current at twice the superconducting gap, $V= 2\Delta/e$, giving a direct measure of the gap in the quasi-particle excitation spectrum. \cite{giaever} In the range of voltages below the gap one may observe smaller current steps at $eV=2\Delta/n$, even at very low temperatures. \cite{T&B} This subgap structure was explained by Schrieffer and Wilkins \cite{S&W} in terms of higher order tunnelling processes, with two or more particles crossing the barrier simultaneously (multiple particle tunnelling). When the tunnelling barrier is characterised by a transmission probability $T$, multiple particle processes are expected to scale as $T^n$, for $n$ particles simultaneously crossing the barrier. The experimental tunnelling curves for planar tunnel junctions are only qualitatively consistent with this interpretation, because the barrier is never perfectly homogeneous, \cite{foden} requiring sums over various $T$'s to the powers $n$.

Although the IV characteristics for tunnel junctions could be described by this mechanism there was the puzzling fact that micro-shorts, or point contact junctions, show similar subgap features. The latter were explained in terms of a model describing the central region of the contact as a normal metal (N) coupled on both sides to bulk superconductors (S), where the electrons in the normal metal undergo Andreev reflection at the NS interfaces. \cite{OBTK} The subgap anomalies result from multiple reflections of electrons, scattering back as holes toward the other interface, reflecting again as electrons, etc. (multiple Andreev reflection). Arnold \cite{arnold} showed that this mechanism can be generalised to any type of junction between two superconductors and specifically to superconducting point contacts. He also included the phase memory of the particles between reflections, which was absent in the models by Klapwijk {\it et al.} \cite{OBTK} The only parameter is the transmission probability $T$ of the barrier, and the microscopic details of the junction are irrelevant, as long as it is short on the scale of the coherence length, $\xi_0$. It is now understood that multiple particle tunnelling is the low-transparency ($T\ll 1$), perturbation theory limit of the more general process of multiple Andreev reflection.

For a quantitative test of the theories a uniform junction with a unique transmission probability is required. This was experimentally realised by forming a single-atom niobium tunnel junction, \cite{niko} which can be described by a single conduction channel on each side of the junction. This reduces the system to an effectively one-dimensional problem, for which several groups worked out an exact description valid over the entire range of transmission probabilities $0\le T\le 1$. \cite{bratus,averin,cuevasSG} It was shown \cite{niko} that the experimentally observed current steps at $eV=2\Delta/n$ scale as $T^n$ for small $T$, as expected, where $T$ is directly obtained from the normal state conductance of the junction, $T=G/G_0$, with $G_0=2e^2/h \simeq (12.9$~k$\Omega)^{-1}$ the conductance quantum.

Scheer and coworkers \cite{scheerPRL} realised that this property can be exploited to determine experimentally the number of conductance channels through atomic size point contacts. It had been shown that the conductance for metallic contacts of atomic size is of the order of the quantum of conductance, \cite{curacao} implying that their conduction properties should be described quantum mechanically in terms of the Landauer formalism, \cite{landauer,buttiker,imry}
$$
G=G_0 \sum_j T_j,
$$
where $T_j$ is the transmission probability for electrons in conductance channel $j$. For a single atom contact for aluminium Scheer {\it et al.} \cite{scheerPRL} found that the number of channels is three, although the conductance is close to one. This was explained using a tight-binding model for the electronic structure of atomic point contacts by Cuevas {\it et al.}, \cite{cuevasTB}, which agrees qualitatively with first principles model calculations. \cite{lang,wan} The general picture that emerged, was that the number of conduction channels through a single atom is determined by the number of atomic valence orbitals. For monovalent metals this number is one, for $s$-$p$ metals three and for $s$-$d$ metals it is generally equal to five. The total conductance depends on the number of electrons occupying the orbitals (the valency) and somewhat on the coupling to the leads, but can be considerably smaller than $NG_0$, with $N$ the number of channels. This interpretation was successfully tested \cite{scheerN} for the metals, gold, aluminium, lead and niobium. Here, a more complete account is given of the experimental results and fitting procedure for niobium. The most important elements are: (1) The observation that the far tunnelling regime can be described by a single dominant conduction channel and that the subgap structure can be reproduced without any adjustable parameters. (2) The number of channels for a single niobium atom is found to be 5, in agreement with the model by Cuevas \ea . (3) For decreasing vacuum tunnelling barrier up to three channels contribute to the tunnel current. (4) An analysis is given of the limits of accuracy of the parameters describing the transmission probabilities in the contact regime. We start by outlining the theory and demonstrating that the current can be expressed as a sum over the individual channel contributions.

\section{Theory}

For the calculation of the current through atomic-size superconducting point contacts, two different approaches have been employed. One approach follows Landauer scattering theory \cite{landauer,buttiker,imry} extended to the junctions with superconducting reservoirs. \cite{bratus,averin} Within this theory, the reservoirs are considered as clean BCS superconductors, and the quasiparticle transmission through the atomic-size contact is described in terms of a scattering matrix. This scattering matrix includes the effect of Andreev reflection by the junction and it is expressed through the scattering matrix of the normal junction.

A different approach was suggested in Ref.~\onlinecite{cuevasSG}, which is based on a tight binding version of the BCS Hamiltonian. Such an approach goes back to the techniques developed by Caroli \ea \cite{caroli} for normal junctions and later applied to resonant tunnelling. \cite{meir} Within this model, the hopping term between the left and right electrodes is treated non-perturbatively, which allows one to eliminate divergences which appeared in early calculations \cite{S&W,hasselberg} based on the tunnel Hamiltonian. The results of calculations of the subharmonic gap structure of the current in single-channel contacts obtained within the improved Hamiltonian approach are essentially the same as the results of the scattering theory calculations. \cite{bratus,averin}

Below, we will sketch the scattering theory method, while a complete account can be found in Refs. \onlinecite{bratus}. The starting point is the model for the junction, which consists of two elements: (i) a constriction modelling the superconducting electrodes, which is smooth on the atomic scale, and (ii) a strong atomic-size scatterer in the neck of the constriction modelling the junction area. The length of the constriction $L$ is smaller than any other length in the problem (dephasing and superconducting coherence lengths, elastic and inelastic mean free paths, etc.). The Hamiltonian for such a model has the form
\begin{equation}
\hat H = \bigg[ \frac{(\hat{{\bf p}} - \sigma_ze{\bf A}({\bf r},t))^2}{2m} +
V({\bf r})+ U({\bf r})+e\varphi({\bf r},t)] -\mu\bigg]\sigma_z 
+ \hat \Delta({\bf r},t),
\label{Hamiltonian}
\end{equation}
where $V({\bf r})$ is the  potential defining the constriction, $U({\bf r})$ is the potential of the scatterer, $\hat \Delta({\bf r},t)$ is the superconducting order parameter:
\begin{equation}
\hat \Delta=
\left(
\begin{array}{cc}
0 & \Delta e^{i\chi/2}\\
\Delta e^{-i\chi/2} & 0
\end{array}\right),
\label{Delta}
\end{equation}
$\sigma_i$ is the Pauli matrix in electron-hole space; the choice of units corresponds to $c=\hbar=1$. By introducing quasiclassical wave functions in the superconducting electrodes,
\begin{equation}
\Psi({\bf r},t)=\sum_{j=1}^{N}\sum_{\pm}\psi_{\perp j}({\bf r}_\perp,x)
{1\over\sqrt{v_j}} e^{\pm i\int p_j dx}e^{i\sigma_z\chi/2}
\psi^\pm_j(x,t),
\label{quasiclassic}
\end{equation}
we will eliminate the potential of the scatterer $U$ from Eq. (\ref{Hamiltonian}) and substitute it by the scattering matrix ${\bf S}$,
\begin{equation}
{\bf S}=\left(
\begin{array}{cc}
{\bf r}& {\bf t}\\
{\bf t} & {\bf r}
\end{array}\right),
\label{S}
\end{equation}
which connects normal electron modes in the left $(L)$ and right ($R$) electrodes. In Eq. (\ref{quasiclassic}), $\psi_{\perp j}$ is the normalized wave function of the transverse mode $j$, $p_j=\sqrt{2m(\mu-E_{\perp j})}$ and $v_j=p_j/m$ are the longitudinal momentum and velocity of the quasiclassical electrons. The quasiclassical wave function $\psi^\pm_j$ obeys the reduced 
Bogoliubov-de Gennes equation inside the electrodes:
\begin{equation}
i\dot\psi^\pm_{j}=(\pm v_j\hat p\sigma_z + \Phi_{}\sigma_z +
v_j p_{s} + \Delta\sigma_x)\psi^\pm_{j}
\label{BdG}
\end{equation}
and the boundary condition at the junction, 
\begin{equation}
\left(\begin{array}{c}
\psi_L^-\\
\psi_R^+
\end{array}\right)
=\left(
\begin{array}{cc}
{\bf r}& {\bf t} e^{i\sigma_z\phi(t)}\\
{\bf t} e^{-i\sigma_z\phi(t)}& {\bf r}
\end{array}\right)
\left(\begin{array}{c}
\psi_L^+\\
\psi_R^-
\end{array}\right)_{x=0}.
\label{match}
\end{equation}
The appearance in Eq. (\ref{BdG}) of the gauge invariant potentials ${\bf
p}_s=\nabla\chi/2-e{\bf A}$ and $\Phi=\dot{\chi}/2+e\varphi$ results from separating out the superconducting phase $\chi$ in Eq. (\ref{quasiclassic}); simultaneously, the phase difference $\phi(t)=\chi_R(0,t)-\chi_L(0,t)$ appears in the boundary condition (\ref{match}). In junctions with a constriction geometry, the potentials $p_s$ and $\Phi$ can be omitted from Eq. (\ref{BdG}), and the deviation of $\Delta$ from the constant bulk value is negligible, due to the rapid spreading out of the current, ${\bf p}_s=\Phi =0$, $\Delta=const$. Thus, Eq. (\ref{BdG}) substantially simplifies and reduces to the BCS eigenstate equation.

Eq. (\ref{match}) is valid under two independent assumptions: (i) negligibly small difference between the normal and superconducting electron wave vectors within the reservoirs, which is consistent with the quasiclassical approximation, and (ii) negligibly small electron-hole dephasing inside the contact area which implies that the energy dispersion of the scattering matrix ${\bf S}$ is small within the interval  of order of $\Delta$ near the Fermi energy. The latter assumption is appropriate for non-resonant atomic-size contacts; a more general boundary condition which includes the effects of the resonance transmission and electron-hole dephasing was derived in Refs. \onlinecite{wendin,johansson}. 

The equation $\Phi=0$ yields the Josephson relation between the phase difference and the applied voltage $V$, $\dot\phi=2eV$. The presence of the time-dependent factor in the boundary condition (\ref{match}) implies that the scattering is inelastic and that the outgoing waves of the scattering states consist of the superposition of eigenstates with energies $E_n=E+neV$, shifted with respect to the energy $E$ of the incoming wave by an integer multiple of $eV$, $-\infty <n< 
\infty$. For example, the forward scattering of quasiparticles incoming from the left is described in the vicinity of the junction ($x<<L$) by the wave function
\begin{equation}
{\psi_{Rj}^+ \choose \psi_{Rj}^- }=
\sum_{n=-\infty}^{\infty} { f_{jn} u_n^+ \choose b_{jn} u_n^-}
e^{i s_n \xi_n x/v_j}e^{-iE_nt},
\label{psi}
\end{equation}
$$
\xi_n=
\left\{
\begin{array}{lr}
\sqrt{E^2_n-\Delta^2}, & |E_n|>\Delta \cr 
i s_n\sqrt{\Delta^2-E^2_n},& |E_n|<\Delta 
\end{array}\right.,
s_n=\mbox{sgn} (E_n).
$$
In Eq. (\ref{psi}), $u_n^\pm$ are (non-normalized) Nambu vectors,
\begin{equation}
u_n^\pm={1\over \sqrt 2}
\left(
\begin{array}{c}
e^{\pm\gamma_n/2}\\
s_n e^{\mp\gamma_n/2}
\end{array}
\right),\;\;\;
e^{\gamma_n}={|E_n|+\xi_n\over\Delta}.
\label{u}
\end{equation}
The scattering amplitudes of different side bands $n$ are connected due to the boundary condition (\ref{match}),
\begin{equation}
{{\bf f}\choose {\bf b}}_{n+1}= {\bf M}_n {{\bf f}\choose {\bf b}}_{n-1},
\;\;\;(n\neq 0),\;\;\;\;\;\;\;
{{\bf f}\choose {\bf b}}_{\pm\infty}=0,
\label{matchbf}
\end{equation}
with the matrix ${\bf M}_n$ having the form  
\begin{equation}
{\bf M}_n= e^{\sigma_z\gamma_{n+1}/2} {\bf T}^{-1}e^{-\sigma_z\gamma_n}
{\bf T}  e^{\sigma_z\gamma_{n-1}/2}. 
\label{M}
\end{equation}
The matrix ${\bf T}$ is the transfer matrix for the normal junction associated with the scattering matrix ${\bf S}$ in Eq. (\ref{S}). 

The matrix ${\bf M}_n$ plays in Eq. (\ref{matchbf}) the role of a transfer matrix for quasiparticle propagation, associated with inelastic scattering, along the side band lattice (energy axis). Inside the energy gap $|E_n|<\Delta$, this matrix obeys the standard transfer matrix equation ${\bf M}_n\sigma_z
{\bf M}_n^\dagger=\sigma_z$ which implies the conservation of the spectral current $K_{n}=|{\bf b}_{n}|^2-|{\bf f}_{n}|^2$.

The dc charge current through the junction can be expressed through the
spectrul currents $K_n$ as follows: \cite{bratus}
\begin{equation}
I= {e\Delta\over\pi}\int^{\infty}_{\Delta} {dE\over \xi}
\sum_{n}[K_{n}\{\gamma_m\} - K_{n}\{-\gamma_m\}]\cosh(\mbox{Re} \gamma_n)
\tanh{E\over 2k_{\rm B}T}. 
\label{I}
\end{equation}
There is an important property of the problem, which reduces it to a single-mode calculation, similar to the normal junction conductance \cite{buttiker} and the dc Josephson current. \cite{beenakker} By including the scattering phases and redefining 
the coefficients ${\bf b}$ and ${\bf f}$ in Eq. (\ref{matchbf}), \cite{bratus} one may express the matrices ${\bf M}_n$, and therefore the spectral currents $K_n$, through the product of transmission matrices ${\bf t}{\bf t}^\dagger$ (cf. Ref. \onlinecite{averin}). Thus, by diagonalizing the matrix ${\bf t}{\bf t}^\dagger$, one may present the total current as the sum over contributions of independent transport modes, each contribution depending on the single junction parameter $T_j$, eigenvalue of the transmission matrix ${\bf t}{\bf t}^\dagger$.

\begin{figure}[!t]
\centerline{\includegraphics[width=.45\textwidth]{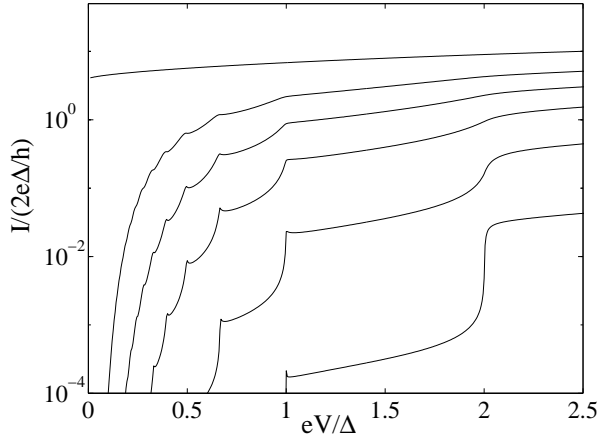}}
\caption{ Current-voltage characteristics of one-channel superconducting junction at at zero temperature; transparency of the junction $T= 1.0,\;0.7,\;0.5,\;0.3,\;0.1,\;0.01,\;$ from top to bottom.} 
\label{fig:theory}
\end{figure}

Equation (\ref{I}) together with the recurrences in Eqs. (\ref{M}) provides a
basis for numerical calculation of the current. 
The calculation of scattering amplitudes in Eq. (\ref{matchbf}), should obey the boundary conditions at $n=\pm\infty$ that the amplitudes approach zero. The simplest way to obtain such solutions is to iterate the recurrences from large $|E_{n}|$ towards $E$. The correct solution will then grow exponentially and numerically `kill' the solution growing at infinity. By this procedure one gets the correct scattering states for each incoming quasiparticle at every energy. The results of numerical calculations of the current-voltage characteristics are presented in Fig. \ref{fig:theory}.

The outlined approach concerns the calculation of the scattering amplitudes of the
superconducting quasiparticles within the electrodes. 
Alternatively, one may separately consider the scattering amplitudes 
of electrons and scattering amplitudes of holes 
by splitting the superconducting quasiparticles into the electron and hole 
components.
One may visualize this splitting by introducing small auxiliary normal 
regions at the edges of the superconducting electrodes and considering the 
electron and hole amplitudes within these normal regions. \cite{averin,hurd,tomas,johansson2} Then the 
inelastic quasiparticle scattering can be formulated in terms of 
electron/hole trajectories which repeatedly traverse the junction and 
undergo multiple Andreev reflections at the SN interfaces between the 
electrodes and auxiliary normal regions. This way of calculation follows 
the original approach to the multiple Andreev reflections (MAR) in SNS
junctions in Ref. \onlinecite{OBTK}.
This version of the scattering theory approach has turned out to be 
particularly useful for the analysis of the structure of the current. 
\cite{johansson2} By connecting the spectral currents $K_n$ with 
probability currents $I_n$ along MAR trajectories and mapping on a 1D wave 
guide problem it allows one
(i) to separate the residual current responsible for the
subharmonic gap structure from the current of thermal excitations; 
(ii) to rigorously prove that the residual 
current only has contributions by the MAR trajectories which cross the energy 
gap; this is particularly useful for calculation of the current in 
highly transmissive junctions when a large number of the side bands is 
excited;  (iii) to consistently distinguish the contributions in the 
residual current of MAR trajectories with a different number of complete Andreev 
reflections, which is equivalent to the presentation of the current as a 
sum over multiparticle currents:
\begin{equation}
I= {e\over 2\pi}\sum_{j=1}^{N}\sum_{n=1}^{n=\infty} n\;
\theta(neV-2\Delta)\int^{neV-\Delta}_{\Delta} dE \;
 I_n(T_j,E)\tanh\left({E\over 2k_BT}\right). 
\label{SGS}
\end{equation}
Due to multiple coherent transitions across the junction, the intensity of the 
$n$-particle current in Eq. (\ref{SGS}) is proportional 
to $T_j^n$, and the $n$-particle current switches on at voltage $eV=2\Delta/n$,  
which explains the step-like features in the 
curves in Fig \ref{fig:theory}.  Always existing as soon as $T_j\neq 1$, these 
features are the fingerprints of the coherent MAR. 
The current spikes at voltages $eV=2\Delta/n$ are connected to the opening 
of $n$-particle transport channels. The spikes  are caused by the
singular quasiparticle density of states (DOS) at the gap edges. These current 
features are particularly sharp, and therefore sensitive to the shape of
the DOS, in the low-transmission limit, i.e., in the tunnelling regime.

\section{Experimental techniques}

Metallic quantum point contacts were fabricated using the Mechanically Controllable Break-junction technique. \cite{curacao,muller92a} The metal to be studied has the form of a notched wire, which is fixed onto an insulated elastic substrate with two drops of epoxy adhesive very close to either side of the notch. The substrate is mounted in a three-point bending configuration between the top of a stacked piezo element and two fixed counter supports. This set-up is mounted inside a vacuum can and cooled down to liquid helium temperatures. Then the substrate is bent by moving the piezo element forward. The bending causes the top surface of the substrate to expand and the wire to break at the notch. 

By breaking the metal, two clean fracture surfaces are exposed, which remain clean due to the cryo-pumping action of the low-temperature vacuum can. This method circumvents the problem of surface contamination of tip and sample in STM experiments, where a UHV chamber with surface preparation and analysis facilities are required to obtain similar conditions. The fracture surfaces can be brought back into contact by relaxing the force on the elastic substrate, while the piezoelectric element is used for fine control. The roughness of the fracture surfaces results in a first contact at one point, and experiments usually give no evidence of multiple contacts. In addition to a clean surface, a second advantage of the method is the high degree of stability of the two electrodes with respect to each other. 

IV characteristics were obtained by standard four-point measurement and current bias for atomic size niobium point contacts, and by voltage bias for the higher resistance tunnel junctions. In order to measure the IV characteristics of a superconducting tunnel junction properly, it is essential that all high frequency disturbances are filtered out as their presence will smear out the detailed features related to Andreev reflection. The filtering used for the 
set-up consisted of ferrite core '$\Pi$' filters at room temperature and copper powder filters at helium bath temperature on all four leads going down to the sample and on the two used for controlling the piezo. All measurements were conducted on Niobium wires of 99.98\% purity at temperatures between 1.4 and 1.6K, far below the superconducting transition temperature of 9.0~K. 

\begin{figure}[!t]
\centerline{\includegraphics[width=.45\textwidth]{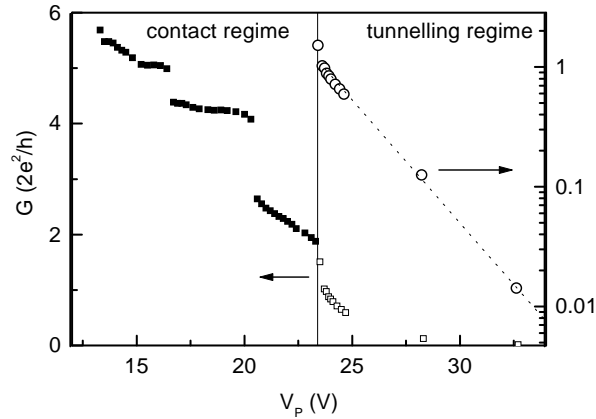}}
\caption{Conductance versus piezo voltage (electrode displacement), where IV curves were taken at each point for analysis of the composition of the transmission in terms of conduction modes. The vertical line divides the IV curves taken in the contact regime (filled squares) from the points taken in the tunnelling regime (open squares).
The points in the tunnelling regime have also been plotted on a logarithmic
conductance scale (right axis, open circles). The calibration of the
piezo voltage is 23 pm/V. The conductance was measured with current bias. Additional points have been taken in the tunnelling regime for a different sample with voltage bias, which are not included in this figure.} 
\label{fig:gplateau}
\end{figure}

Fig.\,\ref{fig:gplateau} shows a curve of conductance against piezo voltage, as an example of the typical behaviour observed for niobium. The measurements were started at a contact consisting of several atoms ($\sim 10 G_{0}$) and each square represents a conductance value at which an IV curve was taken. The solid vertical line in this figure shows the piezovoltage at which the contact breaks, and separates the IV curves recorded in the contact regime from the ones taken in the tunnelling regime. In the contact regime (filled symbols) the conductance shows the characteristic stepwise decrease as a result of atomic structural rearrangements. In the tunnelling regime (open symbols) the points follow the exponential dependence on piezo voltage expected for vacuum tunnelling, as shown by the plot on a logarithmic conductance scale (open circles). The dotted line serves as a guide to the eye, showing a small deviation from exponential dependence close to the transition to contact. We attribute this deviation to the increasing tunnelling contribution through additional channels (see below) and to the forces acting between the front most atoms on the two electrodes, which reduce the actual tunnelling distance from what it is when deduced on the basis of a strict proportionality between piezovoltage and distance. \cite{krans93,olesen96} Although the conductance rises above the quantum conductance,  mechanical contact is only established at the first jump in the conductance, when approaching the transition from the right in Fig.\,\ref{fig:gplateau}. The calibration of the piezo voltage is 23 pm/V ($\pm 20$\%), which is derived from the exponential dependence of the conductance on distance in the tunnelling regime and the bulk value for the work function (3.99\,V). In the following two sections first the IV curves recorded in the tunnelling regime will be discussed before considering the IV curves recorded in the contact regime.

\section{Vacuum tunnelling: parameterless fit of the theory}

\begin{figure}[!t]
\centerline{\includegraphics[width=.45\textwidth]{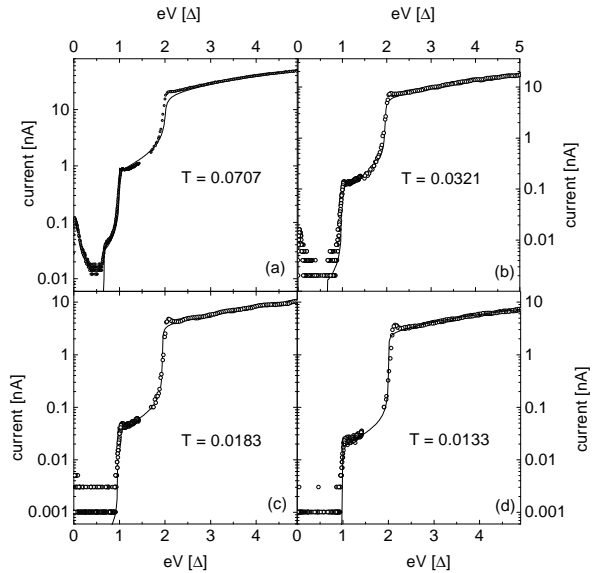}}
\caption{Tunnelling curves for niobium MCB junctions recorded at 1.4\,K. The transmission probabilities have been obtained from the normal state conductance of the junctions assuming a single channel dominates the tunnelling, $T=G/G_0$. The curves have been generated from the theory of Bratus' {\it et al.} \protect\cite{bratus} without any adjustable parameters.}
\label{fig:tunnel}
\end{figure}

Fig.\,\ref{fig:tunnel} shows several examples of IV characteristics recorded in the tunnelling regime, using voltage bias. They have been plotted on a 
semilogarithmic scale to make the steps at small voltages visible. The conductance of the junctions decreases from (a) to (d): $G/G_0=0.0707$, 0.0321, 0.0183 and 0.0133, respectively. The voltage scale is expressed in units of the superconducting gap, which is taken as $\Delta = 1.41$\,meV for these samples. The IV curves have the largest current step at $eV=2\Delta$, as expected for tunnelling, and are linear above this value. Smaller current steps are seen at $2\Delta/2$ and in Fig.\,\ref{fig:tunnel}(a) even at $2\Delta/3$. At still lower currents we are limited by the digital resolution of our experiment, which explains the discrete levels in the figure. The rise in the current for $V$ approaching zero is a remnant of the Josephson current. At $V=0$ we should expect to find the Josephson current due to the coherent tunnelling of Cooper pairs across the junction, and which should ideally have a the 
Ambegaokar-Baratov value $I_c=\pi\Delta/2eR$. \cite{A&B} The latter is of the order of the current just above the step at $eV=2\Delta$ and the observed current is more than two orders of magnitude smaller. This strong suppression and the widening of the Josephson effect into a finite width anomaly can be attributed to the coupling of the junction to its electromagnetic environment, and experiments testing this coupling using a controlled environment are under way. \cite{urbina} The remnant of the supercurrent is more strongly suppressed for weaker Josephson coupling (increasing resistance) as can be seen in Fig.\,\ref{fig:tunnel}, in agreement with Ref.\,\onlinecite{muller92a}. 

The curves through the data in Fig.\,\ref{fig:tunnel} have been generated using the above described theory for a single conductance channel. The only parameters in the theory are the gap $\Delta$, the temperature and the transmission probability $T$. The gap $\Delta=$\,1.41\,meV has been determined experimentally from tunnelling curves with a large vacuum barrier, where subgap structure is absent. The temperature is fixed by the measured bath temperature, and the transmission probability is given by the normal state conductance of the junction through $T=G/G_0$, assuming a single conductance channel. The normal state conductance is determined from the IV curves at bias voltages several times the gap value. This makes the description entirely independent of freely adjustable parameters, and as one can observe in Fig.\,\ref{fig:tunnel}, the theory fits the experiment quite convincingly. The theory reproduces not only the relative height of the current steps at $2\Delta/n$, but also fits the shape of the curve in between the steps. Although the theory incorporates the supercurrent, the coupling to the environment is not included, and we have not attempted to describe this aspect of the experiment.

Apart from the remnant supercurrent feature at low bias, there is a clear deviation at $2\Delta$ in the form of an overshoot in the experimental data compared to the theory. This peak is also seen in planar tunnel junctions for niobium \cite{sherrill} and appears to be a property of niobium. This discrepancy might be explained by the deviation of the real density of states (DOS) in Nb electrodes from the BCS DOS. It was first explained by Wyatt \ea \cite{wyatt} in terms of an atomic-size normal metal layer at the surface of the superconductor. A theoretical study of quasiparticle tunnelling in diffusive proximity SNIS structures showed that the current bump in the vicinity of the $eV=2\Delta$ may result from the suppression of the order parameter at the NS interface. \cite{golubov,ovchin} Spatial inhomogeneity of the order parameter induced by a tiny layer of normal metal causes a deviation of the quasiparticle DOS at the gap edges from the BCS DOS, which gives rise to an enhanced single particle current near the threshold $eV=2\Delta$. It has been shown that such an enhancement is pronounced even for a small deviation of the gap from the bulk value. >From the shape of the curves for our junctions we estimate the thickness of the surface layer at $\sim 1$~nm. 

A model of the junction as two short 1D normal metal wires, proximity-coupled to 1D superconductors, qualitatively reproduces the shape of the tunnelling curves above the gap. \cite{cuevasUP} These deviations from standard BCS behaviour mostly affect the shape of the current steps around $2\Delta /n$ but not the current values in between and the effect becomes less pronounced with increasing transmission probability. An attempt to fit the conductance curves in the contact regime with the modified model of Cuevas {\it et al.} \cite{cuevasUP} did not significantly improve the quality of the fits, at the cost of introducing additional parameters. In the analysis of the curves discussed below we use the standard BCS form, but a low weight is given to the points around $1.95\Delta  <V<2.25\Delta $ as in this regime the measured subgap structure deviates most strongly from the calculated BCS current.

The results show that the theoretical description is very accurate, independent on any adjustable parameters. The single channel tunnelling regime clearly illustrates the principle and can be qualitatively understood in terms of distinct $n$-particle processes, having a probability $T^n$. The fit for $T=0.0707$ in Fig.\,\ref{fig:tunnel}(a) is least satisfactory, disregarding, the above mentioned features. This is due to the fact that at close proximity other channels become important, and a single channel description breaks down, as will be discussed in the next section. The interpretation of multiple channels contributing to the tunnelling behaviour is consistent with the fact that the conductance rises above the quantum unit, while the junction has still a vacuum barrier (Fig.\,\ref{fig:gplateau}), as is evident from the smooth, nearly exponential dependence of the conductance with distance. The first point in the tunnelregime in Fig.\,\ref{fig:gplateau}, at $V_p$=23.5\,V, is well described by three channels. As the vacuum tunnelling gap is increased gradually a single channel remains for conductances larger than approximately $0.05G_0$. As will be shown, a single atom niobium contact admits five channels, composed of one 5$s$ and five 4$d$ orbitals. The reason why a single channel survives in tunnelling, is that the exponential decay of the wave functions selects the one which extends farthest into the vacuum. This wave function is probably associated with the 5$s$ orbital.

\section{Single atom contacts}

\begin{figure}[!t]
\centerline{\includegraphics[width=.45\textwidth]{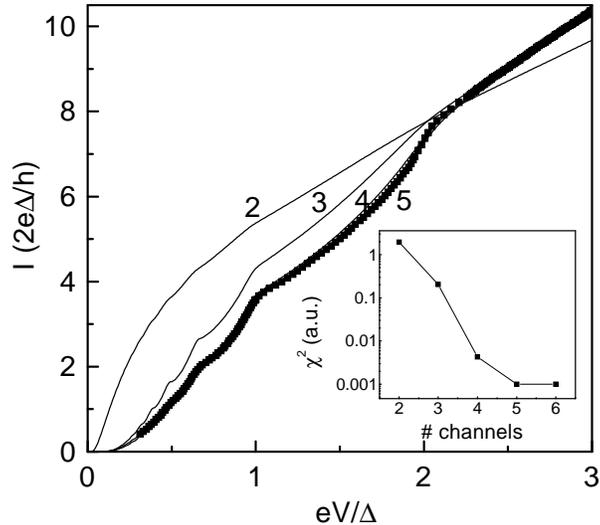}}
\caption{IV curve measured on the first point of the last plateau, at $V_p$=20.6\,V in Fig.\,\protect\ref{fig:gplateau} (solid squares), together with best fits for two, three, four and five channels. The theoretical curves have
been labelled corresponding to the number of channels used in the fit. For the  gap $\Delta/e$=1.41\,mV was used taken from the vacuum tunnelling data. The five channel fit has transmission probabilities $T_1$=0.811, $T_2$=0.669, $T_3$=0.627, $T_4$=0.327 and $T_5$=0.131.  The total transmission probability is $T_{\rm tot}=\sum_{j=1}^5 T_j=2.57$. The inset shows the error sum, $\chi^{2}$, versus the
number of channels used in the fits. }
\label{fig:5channels}
\end{figure}

In the contact regime, shown in the left half of Fig.\,\ref{fig:gplateau}, the conductance evolves in a series of steps and plateaux as the electrodes are pulled apart by increasing the piezovoltage. The last plateau, with a conductance between 2 and 3\,$G_{0}$ is expected to consist of a single niobium atom. At each point IV curves were recorded and as an example we show in Fig.\,\ref{fig:5channels} a curve recorded at the first point of the last plateau. Following Scheer \ea \cite{scheerPRL,scheerN} we assume that the total current can be decomposed into the sum of the individual channels contributing to the conductance, 
$$
I(V)=\sum_{j=1}^{N}i(T_{j},V), 
$$
where $i(T_{j},V)$ is the current of channel $j$. This procedure is justified, as discussed in Section II, since multiple Andreev reflections do not mix the normal conduction channels. This property allows us to extract the mode composition of atomic size contacts from its IV characteristic, using the exact single channel functions $i(T_{j},V)$ for BCS superconductors calculated in Section II.
 
The recorded IV characteristic can be well described by the five channels with transmission probabilities listed in the figure. For comparison also fits using four, three and two channels are shown. Using four channels to fit the data results in a slightly lower quality fit, where it overestimates the current at voltages between 
$\Delta $ and $2\Delta $, and underestimates below $\Delta $. The quality of the fit is expressed in a so called $\chi ^{2}$ factor, which is a the sum of the square of the deviation of the measured current at all the points on the experimental curve from the current of the calculated curve at the same bias voltage, divided by the number of recorded points. Using three or two channels to reproduce the experimental data results in increasingly poorer fits and consequently in larger values for $\chi ^{2}$ (inset). Using six channels to fit the data does not result in any improvement with respect to five channels, and the value of the sixth channels remains near its insignificantly small initialisation value ($T_{6}=0.0001$). Hence, it can be concluded that five channels are necessary to reproduce the IV characteristics measured on the last plateau before breaking the contact. 

All curves recorded at the last conductance plateau, i.e. for $V_p$ between 20.6 and 23.3\,V in Fig.\,\ref{fig:gplateau}, are well described by five channels. For the IV curves recorded at larger contacts ($G>3G_{0}$) it can only be stated that more than seven channels are required to reproduce the measured data with a theoretical curve. The IV curve in this regime can only hesitantly be related to a definite number of channels, as with this large amount of parameters it becomes difficult to decide whether one channel more, really produces a better fit of the experimental data. It is stressed however, that a fit with less than seven channels in all cases produces an unsatisfactory result. 

\begin{figure}[!t]
\centerline{\includegraphics[width=.45\textwidth]{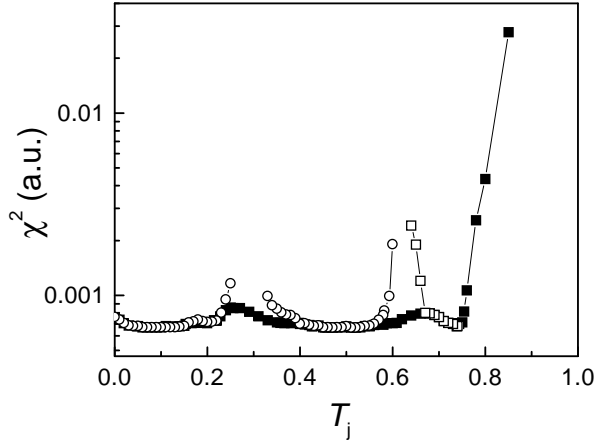}}
\caption{Plot of the quality of the fit ($\chi^{2}$) while fixing one transmission value ($T_{1}$) versus the value at which it is fixed (solid squares). The open squares show the result for $\chi^2$ when all $T_j$ are coerced to lie below the given $T$ value. This procedure shows that one of the $T_j$ has a value between 0.67 and 0.75. In a next step we fix this $T_{1}$ at 0.74, and fix a second transmission probability at various lower values, where we optimise the other three $T_j$ and evaluate $\chi^2$ (open circles).}
\label{fig:fit1fixed}
\end{figure}

The actual values of the transmission probabilities obtained from the fits such as the ones listed in in Fig.\,\ref{fig:5channels}, should be interpreted with care. An estimate for the error in a fit parameter can be obtained by simply changing one transmission probability slightly, recalculating the resulting theoretical IV curve and calculating the quality of the fit, $\chi^{2}$. This procedure, however, will give us an overestimate of the accuracy for this parameter. A better method involves a re-minimisation of the other parameters after changing the one for which the error is being determined. The measured IV characteristic for the last point before the jump to tunnelling, at $V_p$=23.3\,V, has been studied by performing a five channel fit while fixing  one parameter ($T_{1}$) and repeating this for a number of values $0\le T_1\le 1$ (Fig. \ref{fig:fit1fixed}, solid squares). The results of this fit procedure are somewhat dependent on the starting conditions. This is attributed to the presence of local minima in which the fit gets `trapped'. Each of the points in Fig. \ref{fig:fit1fixed} has been obtained, by testing many starting conditions and taking the lowest $\chi^{2}$ found. Surprisingly, one finds that $\chi^{2}$ is minimal for nearly all values of $T_{1}<0.75$. However, when an attempt is made to fit the experimental curve while forcing $T_{1}$ to have a value larger than 0.75, the quality of the fit rapidly deteriorates. This value can hence be taken as an upper bound for the set of transmission probabilities contributing to the conductance. The lack of pronounced minima in $\chi^{2}$ at a particular set of transmission values indicate that the minima in $\chi^{2}$ for each of the individual channels are broad and overlap with each other. 

The flat character of $\chi^{2}$ with conductance does not prevent us from extracting more information about the channel transmissions. Although it seems that any $T_{1}$ is as good as any other, the remaining four parameters are correlated to this fifth one. What happens with the obtained transmission values while performing the procedure discussed above is that two channels can switch places, but always have definite ranges within which they vary. These ranges can be obtained by further constraining the fit parameters. When a maximum value is coerced on all the five transmission values in the normal fit procedure (Fig. \ref{fig:fit1fixed}, open squares) a slow increase below 0.74, and a sharp rise below 0.66 is observed. From this behaviour of $\chi ^{2}$, it can be concluded that at least one transmission probability must be in this range. To extract the four remaining transmission values, the largest transmission probability was fixed at the lowest value for $\chi^{2}$ in this minimum, $T_{1}=0.74$. In addition, the transmission probability of the second channel was set at a particular value and a re-minimisation for the remaining three channels was performed. The results of this extra step have been plotted as open circles in Fig.\,\ref{fig:fit1fixed}. Two new minima appear between 0.4 and 0.57 and between 0.04 and 0.22, with two channels in each. It therefore can be concluded that there is a single largest channel with $T_{1}=0.74\pm 0.02$ which can be determined quite accurately, and two channels, $T_{2}$ and $T_{3}$, lie in the range 0.34 to 0.57. Note, however, that both values cannot vary over the full range uncorrelated. Roughly, it is estimated that $T_{2}=0.41\pm 0.7$ and $T_{3}=0.50\pm 0.7$. For the other minimum between 0.04 and 0.22 similar arguments apply, as here also the range of possible values of both channels overlap, it is estimated that $T_{4}=0.10\pm 0.07$ and $T_{5}=0.15\pm 0.07$. 

\section{Conductance histogram}

\begin{figure}[!t]
\centerline{\includegraphics[width=.45\textwidth]{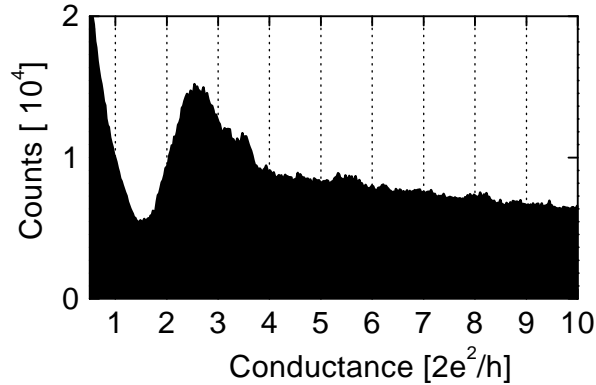}}
\caption{Histogram of conductance values obtained from a set of 4000 curves of conductance vs. piezovoltage, at a temperature of 13~K, well above $T_{\rm c}$.}
\label{fig:histogram}
\end{figure}

From the description of the single-atom conductance in terms of the atomic valence orbitals \cite{cuevasTB,scheerN} the number of conductance channels is well determined, but the total conductance depends on the coupling of the atom to the atoms in the banks. We cannot obtain direct information from the experiment on the atomic arrangement around the central atom. However, the average value of the conductance for an ensemble of contact configurations can be obtained from a histogram of conductance values recorded for a large number of curves of conductance versus piezovoltage. Fig.~\ref{fig:histogram} shows a typical histogram obtained for the normal state conductance, at $T=13$~K. The large divergence at low conductance results from the fact that the jump from contact to tunnelling in Nb is small, and that the conductance in the tunnelling regime initially decreases relatively slowly with distance, starting from values near 1~$G_0$. There is only one pronounced peak in the histogram, centered at 2.6~$G_0$ and about 1~$G_0$ wide. 

Conductance histograms were first introduced to investigate conductance quantization in metallic contacts. \cite{olesen95,krans95} For simple monovalent metals, such as Na and Au, the histograms have been interpreted as being the result of a point contact conductance which is largely determined by quantization. However, for $sp$ and $sd$-metals many channels are available for conduction through a single atom, and these channels are in general only partially open. The result is that peak structure in conductance histograms is not dominated by the quantum nature of the conductance modes, but by the geometrical constraints in forming atomic size contacts. Specifically, the single peak in Fig.~\ref{fig:histogram} should be attributed to the conductance of a single atom. For larger conductances, the many possible geometries and couplings between the atoms prevent appearance of additional peaks. The conductance of $\sim$2.6~$G_0$ for a single Nb atom is in excellent agreement with the values obtained from the theory. \cite{cuevasTB,scheerN}

\section{Conclusions}

From the measurements it can be concluded that the last plateau for niobium with a conductance usually between 2 and 3 $G_{0}$ is composed of 5 channels. This is in excellent agreement with the number of channels and the total conductance predicted by the theory of Cuevas \ea \cite{cuevasTB} using the orbital nature of a single atom as a starting point. The individual transmission values usually include a single dominant channel, two medium-sized and two smaller ones. This channel distribution is also in good agreement with the theory, which predicts a single largest channel as a result of the hybridisation between the $s$ and $d_{z^{2}}$ orbitals. The orthogonal combination of these two is closed, and the remaining four channels are distributed over two degenerate sets of channel, with medium-sized and small transmission. The agreement between theory and experiment lead us to conclude that the last plateau of niobium with a conductance usually between 2 and 3 $G_{0}$ consists of a single atom, and that the valence orbitals determine the quantum conductance channels through this atom, in agreement with similar observations for Pb, Al and Au. \cite{scheerN}

For contacts with a larger conductance more than five channels are required to properly fit of the experimental data, consistent with the idea that these plateaux correspond to contacts where the narrowest cross section consists of more than one atom.

In the tunnelling regime, just after the contact breaks, nearly exponential behaviour is observed in the conductance while the total transmission probability is larger than one. To describe these tunnel junctions with very small vacuum barriers usually three channels are required. This peculiar behaviour is attributed to a significant overlap of several orbitals when the vacuum barrier is very short. As the electrodes are pulled further apart the transmission decreases exponentially, and the number of contributing channels eventually reduces to a single one.

This work is part of the research program of the Stichting voor Fundamenteel 
Onderzoek der Materie (FOM), which is financially supported by NWO. We thank 
A. Levy Yeyati, A. Mart\'\i n-Rodero, J.C. Cuevas, E. Scheer, N. Agra\"\i t, G. Rubio and C. Urbina  for helpful discussions. Assistance of \AA. Ingerman and J.
Lantz is gratefully acknowledged.

* Present address: Shell International Exploration and Production B.V.,                                            Volmerlaan 8, 2288 GD Rijswijk, The Netherlands. 

\dag  Correspondence should be addressed to JMvR, ruitenbe@Phys.LeidenUniv.nl.


\end{document}